\documentclass[sigconf,screen,authorversion]{acmart}

\usepackage[utf8]{inputenc} 
\usepackage{csquotes} 
\usepackage{multicol}
\usepackage[nameinlink]{cleveref}
\usepackage{array}
\usepackage{makecell}
\usepackage{balance}

\copyrightyear{2020}
\acmYear{2020}
\setcopyright{acmlicensed}
\acmConference[SIGCSE '20]{The 51st ACM Technical Symposium on Computer Science Education}{March 11--14, 2020}{Portland, OR, USA}
\acmBooktitle{The 51st ACM Technical Symposium on Computer Science Education (SIGCSE '20), March 11--14, 2020, Portland, OR, USA}
\acmPrice{15.00}
\acmDOI{10.1145/3328778.3366893}
\acmISBN{978-1-4503-6793-6/20/03}

\begin{document}
\fancyhead{}

\title[Benefits and Pitfalls of Using CTF Games in University Courses]{Benefits and Pitfalls of Using Capture the Flag Games\\in University Courses}

\author{Jan Vykopal}
\orcid{0000-0002-3425-0951}
\affiliation{
  \institution{Masaryk University}
  \country{Czech Republic}
}
\email{vykopal@ics.muni.cz}
\authornote{A part of this work was done during a stay at National University of Singapore as a research fellow in National Cybersecurity R\&D Laboratory.}

\author{Valdemar Švábenský}
\orcid{0000-0001-8546-280X}
\affiliation{
  \institution{Masaryk University}
  \country{Czech Republic}
}
\email{svabensky@ics.muni.cz}

\author{Ee-Chien Chang}
\orcid{0000-0003-4613-0866}
\affiliation{
  \institution{National University of Singapore}
  \country{Singapore}
}
\email{changec@comp.nus.edu.sg}

\begin{abstract}

The concept of Capture the Flag (CTF) games for practicing cybersecurity skills is widespread in informal educational settings and leisure-time competitions. However, it is not much used in university courses. 
This paper summarizes our experience from using jeopardy CTF games as homework assignments in an introductory undergraduate course. Our analysis of data describing students' in-game actions and course performance revealed four aspects that should be addressed in the design of CTF tasks: scoring, scaffolding, plagiarism, and learning analytics capabilities of the used CTF platform. 
The paper addresses these aspects by sharing our recommendations.
We believe that these recommendations are useful for cybersecurity instructors who consider using CTF games for assessment in university courses and developers of CTF game frameworks. 

\end{abstract}



\maketitle

\section{Introduction}

Capture the Flag (CTF) games are widely used for cybersecurity competitions and awareness events~\cite{ctftime}. Teams of players solve several problems of varying complexity in a limited time ranging from hours to days. Assignments 
usually contain only a file or the IP address of a system that has to be analyzed and very little or unclear instructions on what to do. The only clear goal is to find the flag (a string). Figuring out a feasible approach and clues provided by the file or system is considered a part of the game.  Players can submit (un)limited number of attempts without any penalty and immediately see whether their submission is correct as well as the score of their competitors. Some games also provide hints, which may cost some points the players already earned in the game.

On the contrary, summative assessment in university courses is usually less interactive and dynamic. Final exams, midterm tests, and homework assignments are traditionally completed individually rather than in a group. They provide more information in the assignment, and students must submit more evidence than just one string in a time range of several hours (exams and tests) or weeks (homework).
In addition, only one submission is accepted, and students have to wait several days or weeks for its marking. Hints might be provided only for homework. 

From the comparison above, CTF games seem to be a better assessment method of skills acquired during the semester, especially for large classes. Gamification features should bring students a more enjoyable learning experience, including not only technical skills but also teamwork. Instructors should benefit from automatic scoring of students' submissions and spend time consumed by the manual marking of students' submissions more efficiently.

However, there is not much research literature investigating the suitability of the use of CTF games in university courses. This paper fills this gap by discussing the results of a case study of using jeopardy CTF games as homework assignments in an introductory computer security course taught at a public university. We are interested in how students apply taught skills and knowledge in CTF games and what are the advantages and drawbacks of using such gamification in the context of tertiary education. We share our experience and provide recommendations for instructors and developers of CTF platforms. We also contribute to the field by developing two open-source software plugins for a popular jeopardy CTF platform~\cite{chung2017}.

\section{Related work}

Although it is generally accepted that CTF games are an engaging and popular education tool, only a few works in the context of tertiary education can be found in the literature. 

\subsection{CTF in competitions} 

The original purpose of CTF was competitive~\cite{chung2017}, and most CTFs remain highly focused on competition~\cite{taylor2017}. Similarly to programming contests~\cite{kirkpatrick2016}, their goal is to showcase and evaluate the performance of already skilled participants~\cite{feng-191757}. Competitive CTFs cover many cybersecurity topics and offer recruitment opportunities and reputation building~\cite{pusey2016}
to the participants. Next, a competitive setting can motivate and engage students, especially those who are attracted to cybersecurity, have extensive prior experience, or possess skills required by the competition~\cite{tobey2014}. 
By solving the competition tasks, participants deepen their understanding of cybersecurity~\cite{weiss2016balance} and practice creative approaches to both known problems and those outside the traditional curriculum~\cite{chapman2014}. 
Moreover, competitions offer considerable learning benefits also before and after the event. Preparing for a CTF involves developing new tools, studying vulnerabilities, and discussing strategies~\cite{vigna2014}, which exposes participants to new skills~\cite{eagle2013}. After a CTF, the competitors or organizers publish \textit{writeups}: walkthroughs that report solutions and explain the vulnerabilities involved in the game. Both writing and reading these is beneficial~\cite{trickel2017}.

However, the competitive setting of CTF games might discourage or even alienate some students~\cite{taylor2017, aycock2018}, especially beginners~\cite{tobey2014}, for three main reasons. First, the tasks are usually too difficult for less-experienced participants~\cite{werther2011}. Second, some of the tasks are also intentionally ambiguous, require a lot of guessing or include artificial obstacles to make them harder to solve~\cite{chung2014}. Third, the participants receive limited individual feedback about their progress. They are often unsure if they are on the right track and usually receive only information about whether the submitted flag was correct or wrong~\cite{chung2014}.
Although the unguided progress inherent for competitions suits advanced learners and can lead to creative solutions~\cite{kapur2012}, it is highly ineffective for beginners~\cite{kirschner2006}. Without guidance, novice students miss essential learning goals and take longer to learn a concept~\cite{weiss2016balance}.

\subsection{CTF in university courses}

Class CTFs~\cite{mirkovic2014} are small-scoped competitions that challenge teams of students against each other in realistic attack-defense scenarios. They have been used in an undergraduate course Introduction to Security at University of Southern California in 2013 and 2014. Students were graded based on their contributions to the team. However, no further details about the assessment are provided.

Chothia and Novakovic~\cite{chothia-191767} studied whether CTFs are effective as an assessment tool in academic cybersecurity courses. They showed that the ability of students to acquire flags in CTF-style challenges is highly correlated with their marks from the written submissions for the \emph{same} challenges, and that flag-only marking may lead to more widespread plagiarism.

Leune and Petrilli~\cite{Leune:2017:UCE:3125659.3125686} conducted a study in which ten undergraduate students taking a cybersecurity class were surveyed before and after a two-week-long CTF. The students developed stronger practical skills by participating in the CTF and overwhelmingly enjoyed it. Nonetheless, no relation of performance in CTF and other assessments in the class is reported.

%
The automatic generation of CTF problems for preventing flag plagiarism is discussed in~\cite{burket-191765}. Each player receives a different version of a task, which leads to a unique flag that must be submitted. Still, 14\% of teams participating in a case study submitted at least one shared flag. Since the game lasted 12 days, limiting online flag disclosure was challenging.
%
Similarly to the automatic problem generation, a scaffolded, metamorphic CTF for reverse engineering proposed by Feng~\cite{feng-191757} provides each student with a unique binary for analysis.


%

\section{CTF games and teaching context}

This paper presents experience learned from two jeopardy CTF games that were a part of a computer security course. The course was taught in English at a public university in Singapore in the first semester of the academic year 2018/2019. The 
games were used as homework assignments in an introductory course on information and system security.
We collected game events generated by students using the CTF portal, answers from two surveys, and analyzed students' marks from other forms of summative assessment of the course. The study was approved by the Institutional review board of the university.

\paragraph{Procedure}

At the beginning of the semester, all students of the course were asked to participate in the study by the first author, who was a guest instructor.
Those who agreed to participate did not receive any incentives or reimbursement for taking part in the study. First, they filled in an introductory survey about basic demographic information, including their 
work and study experience. 
Then, they completed individual homework assignments set as CTF games.
Participants' interactions with the CTF portal (e.g., login, submissions of correct and incorrect solutions, or displaying hints) were automatically logged.
Finally, the participants were surveyed about their 
learning experience with CTFs in the course. The marks from other forms of assessment in the course were obtained directly from the course gradebook in the university learning management system. 
All collected data were analyzed to find common patterns and anomalies describing the performance and engagement of various students in the CTF games.

\paragraph{Participants}
Out of 120~students enrolled in the course, 37~students agreed to participate in the study. The introductory survey was completed by 25~students. The median age of the participants was 23 ($\sigma=1.49$). 
Only two participants had played any CTF game before. No participant was a member of any CTF team. Six were employed in a part-time IT-related job.


\subsection{Features of the selected CTF games} \label{sec:ctf_features}

The CTF games have common features determined by using CTFd version 1.2.0, a popular open-source CTF framework~\cite{chung2017}: 
\begin{itemize}
    \item \emph{Challenge value} -- Each problem (question) has a set point value, which is known to students. This value usually reflects the difficulty of the problem.

    \item \emph{Immediate response} -- After a flag (answer) is submitted to the CTF portal, players immediately see whether it is correct or not. They can also submit an unlimited number of incorrect flags with no penalties.
    
    \item \emph{Scoreboard} -- The total current score of all students is available to all students playing the game. 
    Anonymized names of the players were displayed to protect their privacy.

    \item \emph{Scaffolding} -- Some challenges offer one or more hints. 
    Each hint has its set cost: penalty points that will be deducted from the current player's score once the hint is displayed to the player. Hints can also be provided for free (cost 0 points). Hints can be released together with challenge assignment or later during the game, based on current players' progress. 

    \item \emph{Challenge chains} -- By default, there is no hierarchy or dependency of challenges in the CTF portal, so the players can display and solve any challenge of their choice. Although this design choice is suitable for competitions, we feel that some guidance on which challenge should be solved first can be helpful in educational settings. Our team, therefore, developed a plugin for CTFd, which provides a feature of linear unlocking of challenges~\cite{linear-unlocking}. Instructors can then group some challenges to chains with a defined order of challenges that are unlocked to players once they solve the previous challenge in the chain. For instance, challenges to practice SQL injection, which are of increasing difficulty, can be locked in a chain to guide the student to start with the easiest challenge. This approach can be viewed as a means of achieving game balance~\cite{pusey2014}.

\end{itemize}

\subsection{CTF content and parameters}

The course contains two homework assignments (A1 and A2), which were run as CTF games for individuals. A1 consisted of 8 challenges covering topics taught in the first part of the semester: substitution ciphers, hashing, symmetric and asymmetric cryptography, RSA, and cryptanalysis. A2 consisted of 15 challenges on topics of the second part of the semester: network traffic analysis, port knocking, access control, buffer overflow, command injection, format string attack, and SQL injection. 

The difficulty of challenges varied largely -- from a simple execution of one command to a multi-step solution involving binary debugging and writing a helper exploit script. The difficulty of each challenge was indicated by assigning to a respective category (basic, medium, advanced) and its point value (from 5 to 25, median 15). 

A1 contributed by 10\% to the final grade and was due in 26 days, and A2 by 15\% was due in 24 days. Both assignments included optional bonus challenges for those who were interested in exploring the topics in more depth.

The majority of challenges contained hints which cost 0 penalty points. Next, there was one chain of 3 challenges in A1 and two chains of 3 challenges each in A2.

The first assignment also contained optional challenges for familiarization of students with CTF portal and conventions: flag format, hint displaying, and unlocking challenge in the chain.

\section{Experience report} 


\subsection{Students' performance}

We hypothesized that students who are struggling with CTFs games exercising topics taught at lectures and tutorials would struggle in other forms of continuous assessment and at the exam. Therefore, we sought for dependencies between variables collected in the study. The most indicative variables seemed to be the total score from both CTF games and the number of wrong flags submitted by each student. We used Spearman's rank correlation because our dataset is not normally distributed (e.g. one participant submitted extreme numbers of wrong flags in both A1 and A2). To support our hypothesis, we then focused only on students who finally achieved lower than average total marks from all types of course assessment, excluding homework assignments (i.e., exam, midterm quiz, group presentation, and online quiz).

\subsubsection{Observations about all students}


\Cref{tab:corr} shows statistically significant ($p \leq 0.05$)  Spearman's rank correlation coefficients of variables captured in A1 and A2 games. 
The strongest non-obvious positive correlation among variables capturing performance in CTF and other types of assessment are marked red. These correlations were present between 1) the total CTF score of A1 and A2 including bonuses ($Bonuses$) and marks from all other types of assessment in the course ($All$): exam, midterm quiz, group presentation, online quiz ($r = .50, p \leq 0.001$), 2) the total CTF score including bonuses ($Bonuses$) and marks from the midterm ($Midterm$) ($r = .50, p \leq 0.001$), and 3) the total CTF score ($Bonuses$) and the exam ($r = .49, p \leq 0.001$). The strongest non-obvious negative correlation (marked blue) is between the total number of wrong flags 
in A1 and A2 ($Wrong\_flags$) and marks from the midterm ($r = -.47, p \leq 0.05$).


\begin{table}[t]
\footnotesize
\centering
\begin{tabular}{rrrrrrr}
  \hline
               & CTFs & Bonuses & Midterm & Exam  & All   & Wrong flags \\ 
  \hline
  CTFs         & {\color{gray} 1.00} & 0.63    & 0.35    & 0.31  & 0.31  & 0.11 \\ 
  Bonuses      &      &  {\color{gray} 1.00}    & {\color{ACMRed} 0.50}    & {\color{ACMRed} 0.49}  & {\color{ACMRed} 0.50}  & -0.15 \\ 
  Midterm      &      &         &  {\color{gray} 1.00}    & 0.89  & 0.92  & {\color{ACMDarkBlue} -0.47} \\ 
  Exam         &      &         &         &  {\color{gray} 1.00}  & 0.99  & --- \\ 
  All but CTFs &      &         &         &   &  {\color{gray} 1.00}  & --- \\ 
  Wrong flags  &      &         &         &   &     &  {\color{gray} 1.00} \\ 
   \hline
\end{tabular}
\caption{Statistically significant ($p \leq 0.05$) correlation coefficients of selected variables captured in both CTF games}
\label{tab:corr}
\vspace{-8ex}
\end{table}

\subsubsection{Observations about low-performing students}

Besides obvious strong correlations, we found interesting strong positive correlations between the 
score from A2 and the A2 game session duration ($r = .61, p \leq 0.005$), and between the score from A2 and time difference between solving the first and last challenge in A2 ($r = .59, p \leq 0.01$). 

\subsubsection{Discussion}

Although we observed some statistically significant correlations between variables mined from CTF games and other forms of the assessment, the correlation coefficients range only from -0.5 to 0.61.  
We believe three factors were affecting these results. First, the total score of A1 and A2 has a skewed distribution. Their medians were the maximum possible scores without bonus challenges (i.e. 100 and 150, respectively) and the standard deviations 8.4 and 31.8, respectively. Second, all the hints were for free. Their usage is not reflected in the score, which remains the same even for students who needed different levels of help.
Third, the used CTF portal did not log important game events. For instance, an event of displaying the challenge by a student can be used to analyze how much time the student spent on solving the challenge. However, this event is currently not logged by the CTF portal and cannot be easily and reliably reconstructed from other sources such as webserver access logs. This negatively affects the feasibility or accuracy of some variables relying on the missing event. The manual reconstruction of these events from the logs revealed the strongest non-obvious correlation of any two variables of our dataset.

\subsubsection{Recommendation for instructors}

\paragraph{Examine what kind of game events are logged by the CTF portal}
In our case, both assignments were open for almost a month, which might eventually enable the vast majority of students to complete all challenges with the full score regardless of how much time they needed. In addition, the instructors did not want to discourage students from displaying hints by deducting penalty points. Therefore, displaying these \enquote{free hints} did not affect the variance of the total score.
We recommend focusing on logging features of a CTF platform, in particular, more detailed information about the game progress, such as the duration of students' interactions with the platform. This will deliver more accurate metrics than the points awarded for the submission of correct flags.

\subsection{Usefulness of hints} \label{sec:hints}

The majority of challenges was equipped with one or more hints. We analyzed how they helped students with solving the challenges. 

\subsubsection{Observations}

\begin{figure}[t]
    \centering
    \includegraphics[width=1\linewidth]{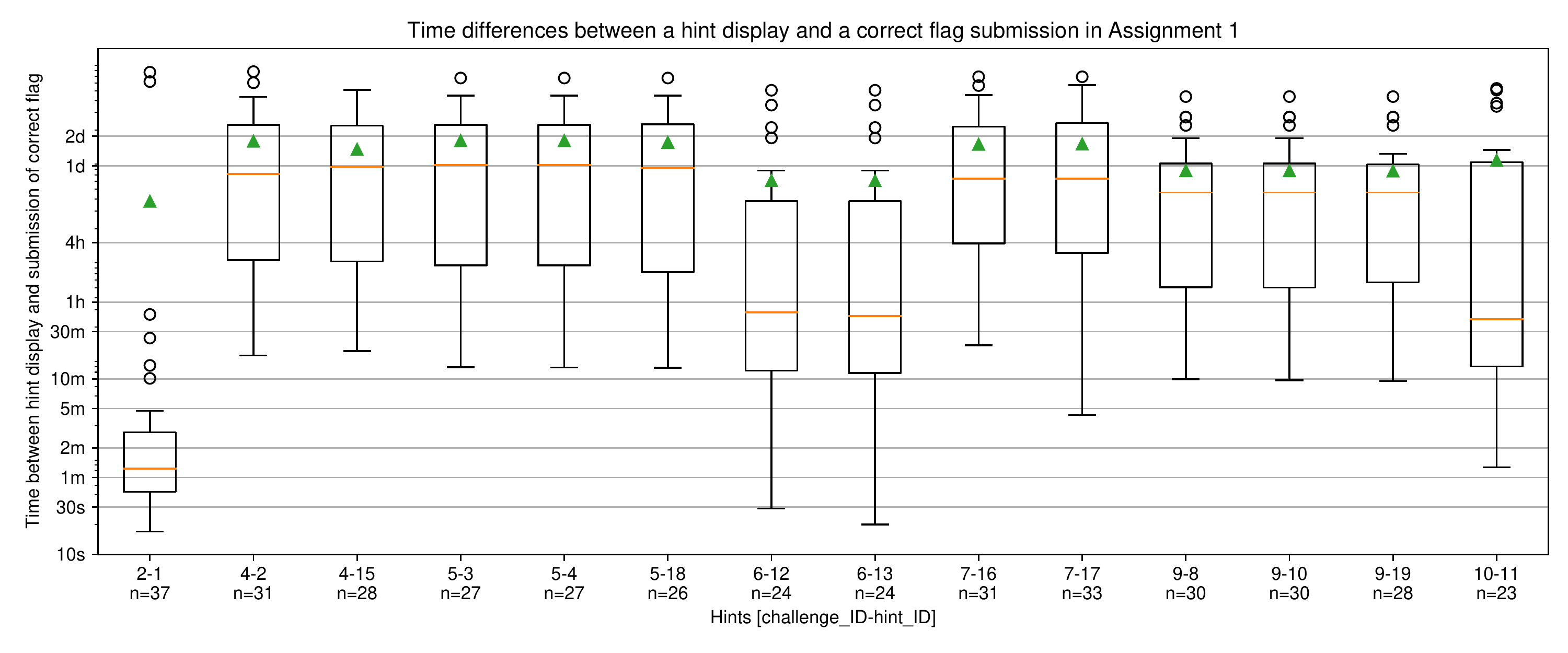}
    \caption{Time differences between a hint display and a correct flag submission in A1. Green triangles denote means.}
    \label{fig:solves_hint_displays-A1}
    \vspace{-4ex}
\end{figure}

\begin{figure*}[t]
    \centering
    \includegraphics[width=1\linewidth]{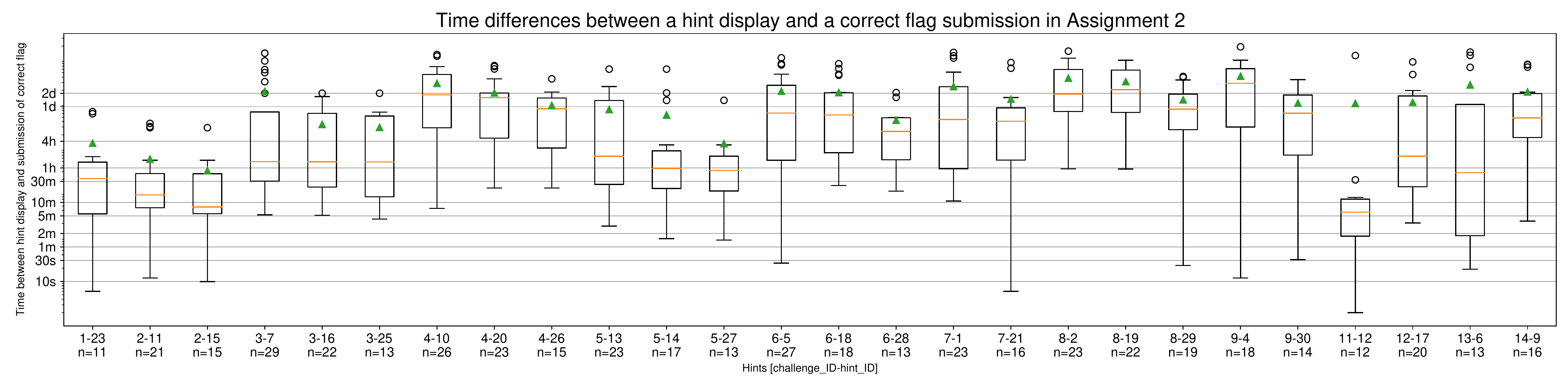}
    \caption{Time differences between a hint display and a correct flag submission in A2. Green triangles denote means.}
    \label{fig:solves_hint_displays-A2}
\end{figure*}

\Cref{fig:solves_hint_displays-A1} and \Cref{fig:solves_hint_displays-A2} show times between displaying hints and solving the respective challenge. 
Only the hints used more than 10 times are plotted. In both games, there were challenges that the students did not solve in a reasonable time despite viewing hints. High median times of 1~day or more can be observed in both figures. In contrast, only 7 challenges out of 19 were solved with the median time less than 1 hour after displaying the hint.

The answers in the after-game survey suggest that some hints did not reach their aim. Only one student assessed hints as \emph{Very much useful}, two as \emph{Much useful}, 8 as \emph{Moderately useful}, 4 as \emph{Slightly useful} and 1 as \emph{Not useful at all}.
In addition, 6 students out of 20 mentioned hints in their answer to the  question \enquote{If you could choose one element of your CTF experience to improve, what would it be?}

\subsubsection{Discussion}

The major differences among median times of individual hints indicate some hints were more helpful than others. This can be illustrated by hint 1 of challenge 2 in A1. This challenge was, in fact, straightforward instruction on how to display and use hints. The median time was the lowest (only 1 minute). 
The instant feedback provided by students on hints right after solving the challenge in the CTF portal supports that some hints were more useful than others. One example is challenge 2 in A2: all 7 students who provided feedback assessed hints as \emph{Rather Useful}, \emph{Somewhat Useful} or \emph{Useful}. The challenge was easy, and hints were clear and guiding, e.g., \emph{You need to view the \_image file\_ to get the flag.}

\subsubsection{Recommendations for instructors}
\paragraph{Indicate what a hint is about} The information about hint cost (if any) is not enough for students to decide whether they will benefit from displaying the hint. We recommend adding a short description of what they can expect, such as \enquote{what tool to use}, particularly in challenges offering two or more hints. Otherwise, students may display a hint which tells them what they already figured out. 
\paragraph{Test challenge assignments and hints before the game} Ask teaching assistants or peer instructors to test the challenge descriptions and hints to balance what should be placed in the challenge assignments and what can be left for one or more hints. While the challenge assignments can be a bit fuzzy, hints should be clear and straightforward.
\paragraph{Prepare backup hints} Although hints have been tested, students may still struggle with some (advanced) challenges. Monitor the ongoing game (submissions, wrong flags, and hint usage) and be ready to add a new hint if needed.

\subsubsection{Recommendations for developers of CTF frameworks} 
\paragraph{Actively offer hints} In our experience, some students tend to beat the challenge without displaying any hints even though the hint may speed up their progress. Consider adding a feature which will offer a hint to a student after some time of the challenge solving.
\paragraph{Support adaptive hints}
Any step toward dynamically generated hints would be beneficial. If the CTF framework provides logs capturing game events and even players' behavior in a virtual environment, these data can be used as input to a module serving the appropriate hint at the right time.

\subsection{Flag sharing in individual CTF}

We investigated four unusual patterns indicative of plagiarism.

\subsubsection{Observations}

\begin{figure}[h]
    \centering
    \includegraphics[width=1\linewidth]{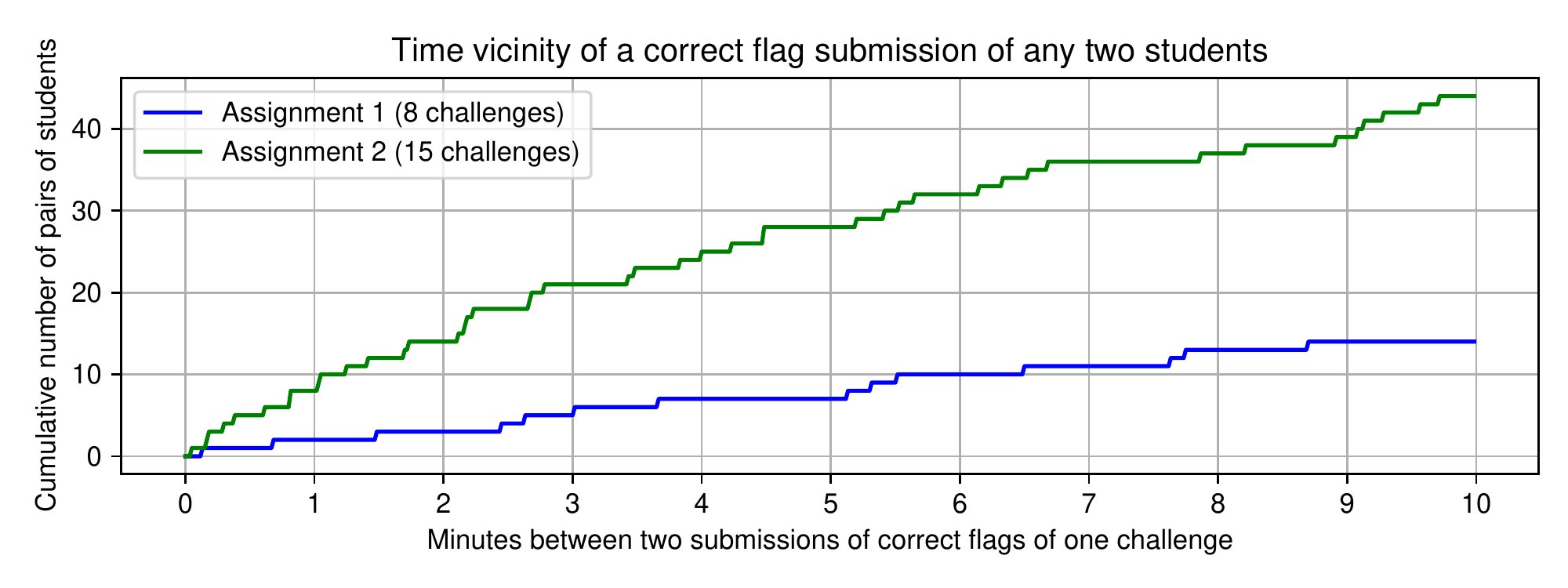}
    \caption{Cumulative number of pairs of students who both submitted correct flag for the same challenge within 10 minutes.}
    \label{fig:solves_time_vicinity}
    \vspace{-2ex}    
\end{figure}

\paragraph{Submissions of the same flag in a time vicinity}
We supposed that if students are solving challenges individually in a time frame of almost a month, it is unlikely that they will submit the correct flag (almost) at the same time. 
\Cref{fig:solves_time_vicinity} presents how many pairs of students submitted the correct flag for the same challenge shortly after each other.  
Two challenges were solved by 8 and 7 pairs of students, respectively, and the rest of the challenges by 4 and fewer pairs.

\paragraph{Correct flag submitted as an incorrect flag to another challenge}
We observed 8 cases in total. 
The time between the submission of any valid flag of other challenges as the wrong flag varied from 14 seconds to almost 18 hours. Notably, there were two cases in A2 where a flag from a still locked challenge was submitted as an incorrect flag to the preceding challenge.

\paragraph{Challenges solved without downloading the file to analysis}
Some challenges require students to visit online services running at other machines than the CTF portal. Others contain attachments that have to be downloaded first. For this analysis, we inspected logs of the web server hosting the CTF portal. 
In total, seven challenges contain 12 attached files required for solving the challenges. 
We discovered 11 cases of solving any such challenge by 6 different students without prior download of a required file. 

\begin{figure}
    \centering
    \includegraphics[width=0.95\linewidth]{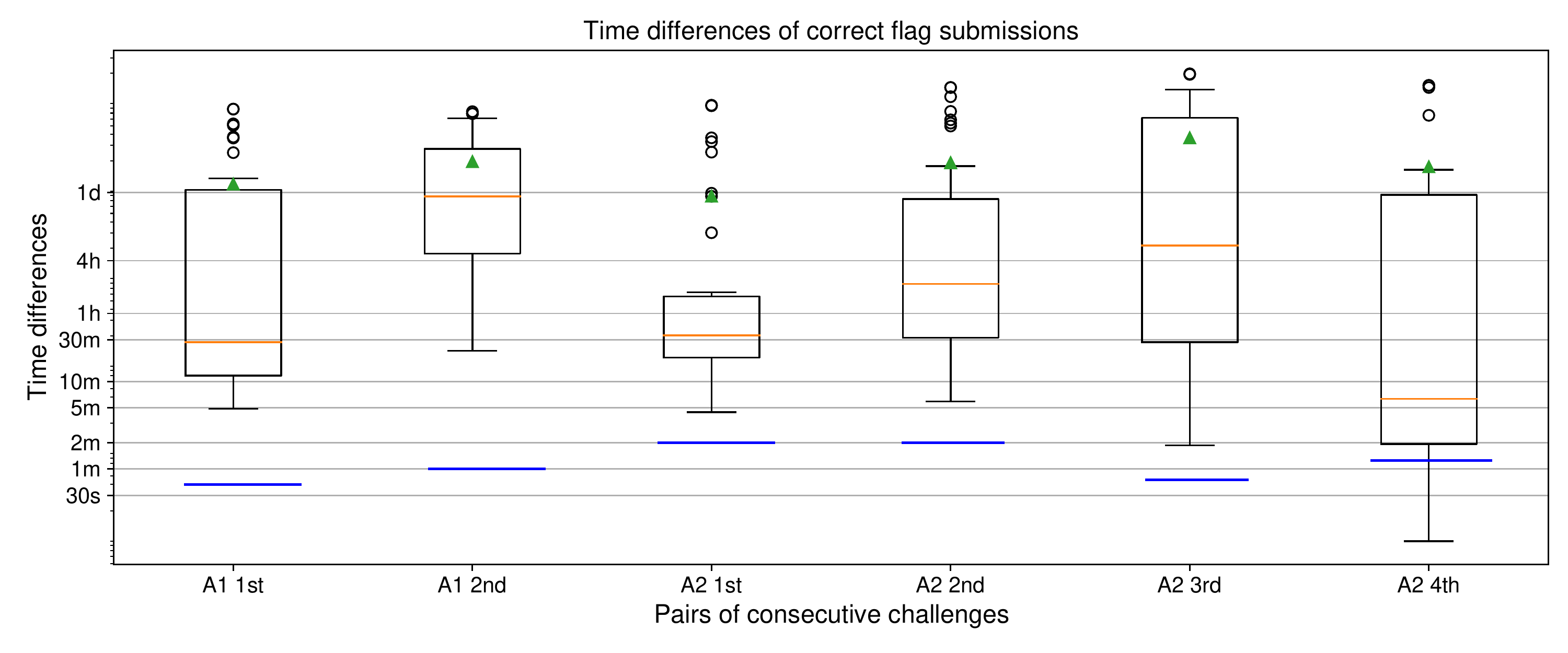}
    \caption{Time between submission of correct flags of two consecutive (locked) challenges. Blue lines indicate the minimal possible solve time of the first challenge. 
    }
    \label{fig:locked-pairs-times}
    \vspace{-4ex}
\end{figure}

\paragraph{Quick solves of consecutive challenges} 
Finally, we focused on linearly locked challenges in both assignments. We estimated the minimal possible solve time by an expert player who immediately figures out all steps required for solving a challenge, performs these steps without any mistakes, and submits the correct flag to the portal. The minimal possible solve times range from 20~seconds to 2~minutes. \Cref{fig:locked-pairs-times} depicts the time differences of the submissions of the correct flag of a locked challenge and a preceding challenge in the same chain. A1 contains one chain with 3 challenges (two pairs of consecutive challenges) and A2 two chains with 3 challenges each (four pairs of consecutive challenges). We discovered 7 very quick solves ranging from 9 to 53 seconds in the fourth pair of challenges in A2 with the minimal possible solve time of 1 minute and 15 seconds.

\subsubsection{Discussion}

These different views on game events revealed some students might have obtained flags unexpectedly quickly, without downloading the necessary file(s), had submitted flags almost at the same time, or had submitted the correct flag to another challenge. While the time vicinity of submissions may report only weak indications of plagiarism, others can be considered as more serious pieces of evidence (such as quick solves of consecutive locked challenges).
In A2, instructors questioned such students. Three of them eventually confessed they used flags shared by their peers. Some students argued this was only a coincidence since they consulted their approach in a group and then solved the challenge and submitted the flag each on their own.

\subsubsection{Recommendations for instructors}

\paragraph{Set rules for students' collaboration during the game in advance} Decide what will constitute plagiarism in your class. Is any discussion about challenges among students forbidden, or do you allow non-detailed discussions about challenge principles or techniques that can be used? Communicate these rules clearly and explicitly to students.

\paragraph{Inform students about how you will check suspicious submissions in advance} 
Describe a procedure that will be applied if instructors spot suspicious behavior. For instance, the instructor may (randomly) select several students for in-person demonstration of how they solved particular challenges.

\paragraph{Structure related problems to challenge chains}
Challenge chains help not only with revealing plagiarism but also explicitly guide students what must be solved first. 

\subsubsection{Recommendations for developers of CTF frameworks}

\paragraph{Support challenge chains or dependencies} 
If your platform does not have this feature, add it.

\paragraph{Provide built-in analyses for revealing flag sharing}
All analyses we performed were done outside the CTF platform using ad-hoc scripts and third-party tools. However, these analyses work only with generic game events such as time and date of flag submissions so that they can be run automatically for any CTF game. If the results of these analyses were easily accessible in the CTF portal, instructors would definitely benefit from them.

\subsection{Students' perceptions of the CTF games}

One week after the games were over, we asked students for their voluntary feedback in the 
online survey. \Cref{tab:questions} lists the questions.

\begin{table}[ht]
    \centering
    \footnotesize
    \begin{tabular}[t]{cl}
      \toprule
      \bf{No.} & \bf{Question} \\
      \midrule
        Q1 & \makecell[tl]{Would you rather complete the Capture the Flag games or normal homework \\ assignments in your future security courses?} \\
      \hline
        Q2 & How useful did you find the hints? \\
        Q3 & \makecell[tl]{How useful was the instant feedback on your submissions \\ (a response of the CTF server whether your flag is correct or not)?} \\
        Q4 & How useful was seeing your overall progress in assignments? \\
        Q5 & How motivating was the scoreboard? \\
        Q6 & \makecell[tl]{How frequently have you talked about your score with your peers \\ during the semester?} \\
        Q7 & How much did you like the unlocking feature? \\
    \bottomrule
    \end{tabular}
    \caption{Wording of questions in the after-game survey.}
    \label{tab:questions}
    \vspace{-8ex}
\end{table}

\subsubsection{Observations}

The crucial question was the first one.
The vast majority of students (13 out of 16) would prefer CTF games, only two students normal assignments and one student was not sure. Students who would prefer the games liked that the games were fun, hands-on, more interactive, objective, allow to learn to work with security tools, and allow lots of trial and error in exploration. The two students who would prefer the normal assignments considered the games difficult and very time-consuming. One of them felt he had not learned much from them. The student who was not sure mentioned that \enquote{the normal assignment is little harder to simply copy than CTF games.}

\begin{figure}
    \centering
    \includegraphics[width=0.95\linewidth]{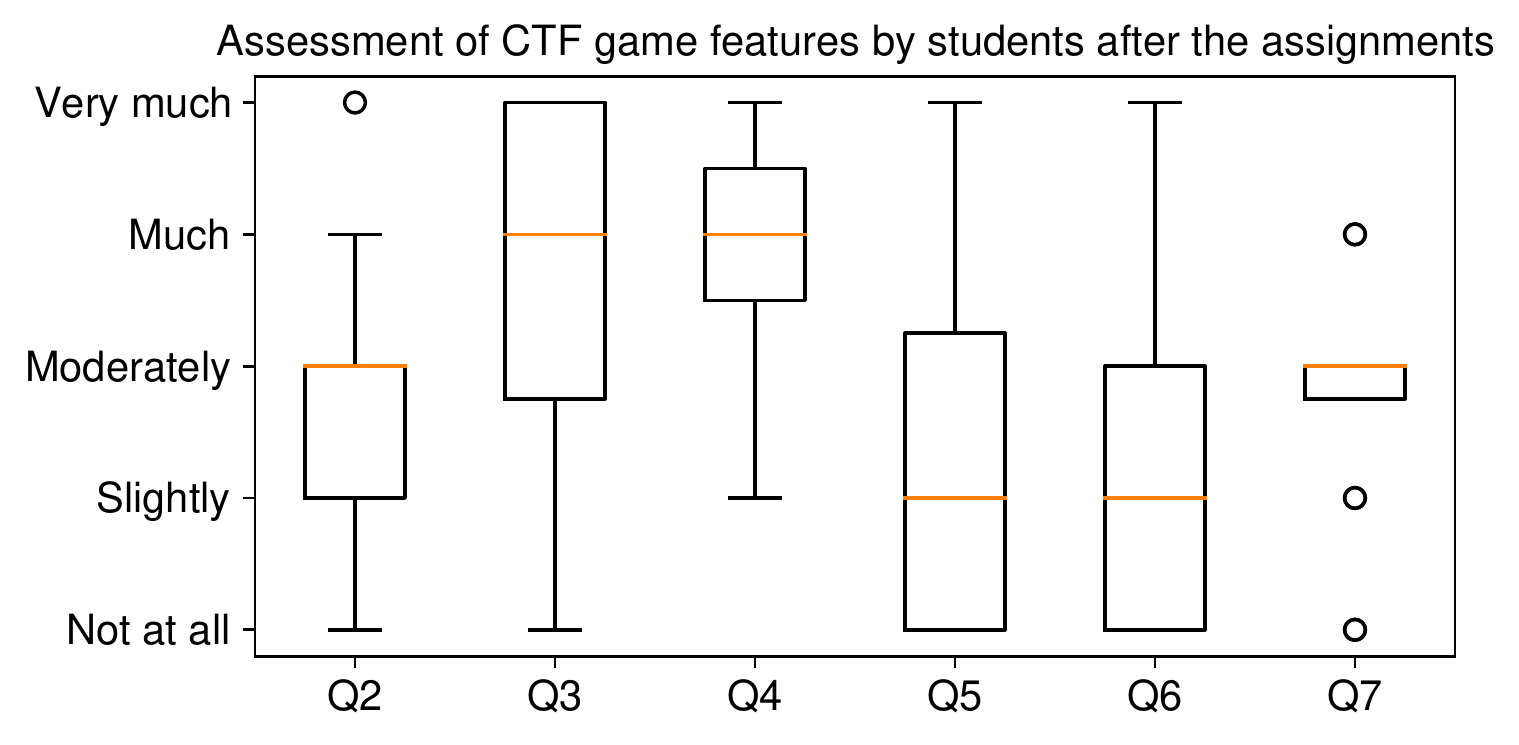}
    \caption{After-game students' assessment of CTF features.}
    \label{fig:final-survey}
\end{figure}

The answers to the other questions (Q2--Q7) are depicted in \Cref{fig:final-survey}, which shows students' assessment of various game features on a 5-point Likert scale. Students could elaborate on their answers to questions 2, 5, and 7.
Student comments' of answers to questions 2 was already discussed in \Cref{sec:hints}. Regarding the scoreboard (Q5), some students would be more motivated if the scoreboard listed real (non-anonymized) names of their peers. Finally, the challenge unlocking (Q7) was used appropriately since it allows to \enquote{solve simpler challenges first and used the thought process and methods to help solve the harder challenges.}

\subsubsection{Limitations}


The final after-game survey was answered by 16 out of 37 students. This could introduce some bias since the rest of the students who did not provide any input may not have liked their CTF experience.

\section{Conclusions}

Capture the Flag games can be an enjoyable way of assessment in university computer security courses. 
We studied the use of jeopardy CTF games in the summative assessment. Based on this experience, we provided recommendations for both instructors and developers of CTF platforms. 

\vspace{-1ex}
\subsection{Summary of our experience}

We conclude that replacing traditional homework assignments by CTF games is generally favorable for both instructors and students. The instructors can save time spent on marking the students' submissions and enable students learning practical skills in an interactive and enjoyable way. The vast majority of students who answered the after-game survey in our study would rather complete the CTF games than regular homework assignments in their future security courses.


We highlighted several pitfalls of using CTFs in summative assessment. Instructors should carefully consider the game format, scoring (distribution of points of game challenges and hint costs), a CTF platform for running the game, and game duration.

First, although CTFs lasting several weeks provide more opportunities and less stressful environment for exploring and learning the topic than the hours-long CTFs, they are much more vulnerable to flag sharing between students than invigilated intensive CTFs.

Next, the interactive nature of the games, particularly instant feedback about the correctness of a solution, can be utilized for identifying students at risk only if the CTF platform provides advanced analyses of students' progress. The scoreboard is not sufficient in weeks-long games since there is a high chance the vast majority of students will finish almost all tasks, as we witnessed in our study.

Finally, the challenges presented in CTFs 
usually require investigating several dead ends before they are solved. Current CTF platforms, however, provide only static scaffolding by offering hints which do not consider the present performance and experience of a player. This results in the low usefulness of hints offered for medium and hard challenges. Serving dynamically generated hints thus presents an interesting direction for further research.



\subsection{Practical contributions}

This paper reports not only the results of the study but also presents several replicable methods of analysis of game events. 
These methods apply to any CTF game with common basic features regardless of the content of challenges. For example, an excessively high number of submitted wrong flags may indicate some students at risk, or unusually quickly solved challenge may reveal a flaw in challenge design or illicit flag sharing. The only prerequisite for these analyses is the availability and sufficient level of detail of game events logged by CTF platforms. 

Finally, we also contributed to the current practice of running CTF games by developing two open-source software plugins. These extend the popular CTFd platform~\cite{chung2017} by features of linear unlocking of challenges and collecting players' feedback right after solving the challenge. The first plugin~\cite{linear-unlocking} is useful for structuring the challenges and revealing shared flags. The second~\cite{challenge-feedback} serves both CTF organizers and researchers interested in players' thoughts about the solved challenge.
In our future work, we plan to develop more plugins implementing learning analytics methods and other features that will enhance students' learning experience and help instructors to design and run CTFs more efficiently.

\begin{acks}

This research was supported by \grantsponsor{ERDF}{ERDF}{} project \textit{CyberSecurity, CyberCrime and Critical Information Infrastructures Center of Excellence} (No. \grantnum{ERDF}{CZ.02.1.01/0.0/0.0/16\_019/0000822}). We also thank NUS Greyhats for their effort spent on the homework assignments and Heng Le Ong for developing the two CTFd plugins.

\end{acks}

\bibliographystyle{ACM-Reference-Format}
\balance
\bibliography{references}

\end{document}